\documentclass[prd,
twocolumn
,notitlepage
]{revtex4-1}
\usepackage{color}
\usepackage[utf8]{inputenc}
\usepackage{graphicx}
\usepackage{xcolor}
\usepackage{braket}
\usepackage{float}
\usepackage{amssymb}
\usepackage{amsmath}
\usepackage{mhchem}
\usepackage[makeroom]{cancel}
\usepackage{upgreek}
\usepackage{commath}
\usepackage{siunitx}

\usepackage[titletoc]{appendix}

\usepackage{hyperref}
\hypersetup{colorlinks=true, linkcolor=magenta, citecolor=magenta, urlcolor=magenta}

\begin{document}
\title { \textbf{{\large {
%Potential 
Interfacial proximity and interplay between Kondo and short-range magnetic correlations in heterostructures}}}}

\author{Julián Faúndez$^{1}$\href{https://orcid.org/0000-0002-6909-0417}{\includegraphics[scale=0.05]{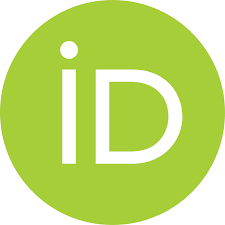}}}

\author{Leonardo C. Prauchner$^{2}$\href{https://orcid.org/0000-0002-8527-7599}{\includegraphics[scale=0.05]{Figures/orcid.png}}}

\author{Peter S. Riseborough$^{1}$\href{https://orcid.org/0000-0002-2216-3586}{\includegraphics[scale=0.05]{Figures/orcid.png}}}

\author{Sebastian E. Reyes-Lillo$^{3}$\href{https://orcid.org/0000-0003-0012-9958}{\includegraphics[scale=0.05]{Figures/orcid.png}} }
 
\author{Sergio G. Magalhaes$^{2}$\href{https://orcid.org/0000-0002-6874-7579}{\includegraphics[scale=0.05]{Figures/orcid.png}}}

\affiliation{$^{1}$Physics Department, Temple University, Philadelphia, Pennsylvania, 19122, USA}
\affiliation{$^{2}$ Instituto de Física, Universidade Federal do Rio Grande do Sul, Porto Alegre, RS,  91501-970, Brazil}
\affiliation{$^{3}$Departamento de Ciencias Físicas, Universidad Andres Bello, Santiago 837-0136, Chile}

\date{\today}

\begin{abstract}

In this work, we investigate the influence of interlayer distance in a heterostructure containing both Kondo effects and short-range magnetic correlations. Our proposed heterostructure comprises three coupled square lattice layers. The first layer is governed by the Kondo-Heisenberg lattice model involving $f$- and $d$-electrons, which interact via Kondo and Heisenberg couplings, $J_{K}$ and $J_{H}$, respectively. The other two layers consist of non-interacting itinerant electrons, where coupling with the first layer is determined by two perpendicular hopping parameters.  We find that varying the interlayer couplings induces electronic dynamics at the interface, altering the behavior of mean-field parameters describing the Kondo effect and short-range magnetic correlations. The system's temperature -
interlayer hopping parameter phase diagram exhibits a sequence of discontinuous and continuous transitions. In the cases,  $|J_{K}|<|J_{H}|$ and $|J_{K}|>|J_{H}|$ rich phase diagrams are found which include Kondo, ferromagnetic and antiferromagnetic correlations. Our work provides insights into hosting Kondo correlations in heterostructures.
%Through the study of both scenarios, similar effects departing from the \textit{half-filling} regime can be observe.

\end{abstract}

\maketitle

%---------------------------------------------------------------------------------------------------------------------------------------------------------
\section{introduction}

In recent years, significant experimental advances have enabled the synthesis of ultrathin layers of heavy fermion electron systems, such as \ce{CeCoIn5} and \ce{CeIn3}, \cite{Naritsuka2019,Zhou2013,Moll2017}. Notably, superlattices or heterostructures can be artificially created by periodically stacking these layers, offering the possibility of modifying the electronic structure and stability of exotic phases. Heavy fermion heterostructures exhibit unique physical properties that are in general distinct from their individual counterparts, including novel states of matter such as magnetism, superconductivity, Kondo effect, the presence of short-range magnetic correlations and classical and quantum multicritical points \cite{Geim2013,Liu2016,Jariwala2017,Lausmann2022,Hyun2013,Bang2020,Willa2021,Faundez2021,Faundez2024}. Overall, the study of such structures provides a means of advancing our understanding of complex electronic systems and developing new technological applications
\cite{Zubko2011,Cao2018,Bistritzer2011,Szeftel2021,Posey2024}.

Interestingly, artificial heterostructures have emerged combining different configurations of lattice geometries and layer materials \cite{Palankovski2004,Kroemer1988,Yan2018,Mazza2021,Yasuda2016,Zhu2021,Bimberg1999}. 
%Artificial heterostructures are engineered materials consisting of thin layers of various compositions, exhibiting a wide range of electronic properties not inherent to any single constituent material. 
In particular, the use of Kondo layers in heterostructures, such as \ce{CeIn3}/\ce{LaIn3}, has resulted in the development of a new class of materials known as Kondo heterostructures. These heterostructures are comprised of alternating layers of Kondo and non-magnetic materials \cite{Shishido2010,Liu2023,Vaňo2021,Huecker2023,Yang2022,Martino2021}.
 
Interface effects can influence the emergence or suppression of several phases, incluinding the normal state ($NS$), the Kondo state ($K$) as well as ferromagnetic correlations ($F_{c}$) and antiferromagnetic correlations ($AF_{c}$) \cite{Liu2023, Euverte2012}. For instance, the presence of interfaces can modulate the electronic density of states in adjacent layers, thereby affecting the formation of Kondo bound states \cite{Endres2002}. Interlayer interactions can perturb local magnetic properties and spin dynamics, consequently suppressing the formation of a Kondo phase \cite{Liu2019, Allerdt2017}. Thus, investigating how the lattice geometry and composition of the heterostructure influence the formation and dynamics of electronic and magnetic states in the presence of interfaces is crucial for understanding Kondo heterostructures \cite{Peters2016,Shi1995,Euverte2012}. Kondo heterostructures have promising applications in the realm of nanoelectronics, including the creation of innovative spintronics devices. Moreover, they can be instrumental in the development of devices for quantum computing. Consequently, the dynamics of Kondo layers within heterostructures represents an intriguing domain within materials physics, offering the potential to unlock a plethora of new electronic functionalities 
 \cite{Peters2016,Yang2019,Naritsuka2021}.
 
A significant number of works have
employed the Kondo lattice model within the fermionic mean-field approximation in a single layer, where ferromagnetic, antiferromagnetic, Kondo and mixed phases beyond the \textit{half-filling} regime have been identified \cite{Iglesiass1997,Ruppental1999,Bernhard2015}. Moreover, studies on a bilayer with a in-situ attractive Hubbard model and Kondo interactions, have lead to the discovery of magnetic, Kondo and superconducting states \cite{Junior2020}. Also, at zero temperature, it has been demonstrated that the competition between the Kondo effect and the RKKY interaction is influenced by the structure of two Kondo layers and a single metallic lattice, where the coupling between layers is given by a homogeneous hopping term $t_{z}$ \cite{Peters2013}. 

Furthermore, a phenomenon known as the \textit{Nozières exhaustion problem} may arise in two-dimensional Kondo lattices. This phenomena occurs when conduction electrons are unable to fully screen all the localized spins, indicating a lower density of conduction electrons compared to localized spins. This issue can lead to the emergence of distinct states of matter, such as magnetic or superconducting phases \cite{Nozieres1998,Coqblin2003}
 
\begin{figure}[H]
    \centering    \includegraphics[scale=1]{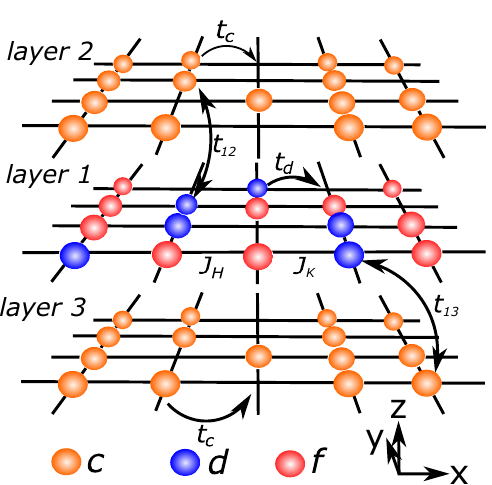}
    \caption{ Illustration of the heterostructure model with three interacting layers.  In \textit{layer 1}, the itinerant part of $d$-electrons is given by the hopping term $t_d$. $J_H$ and $J_K$ describe the Heisenberg and Kondo interactions, respectively. In \textit{layer 2} there are itinerant states with $c$-electrons and hopping term $t_c$ between neighborhoods. The coupling between $c$- and $d$-electrons is determined by the hopping term $t_{12}$.  In \textit{layer 3} the hopping of $c$-electrons is given by $t_c$ and the coupling between $c$- and $d$-electrons is determined by the hopping term $t_{13}$. Furthermore, each layer has an specific bandwidth: $2W^{1}_{d}$, $2W^{2}_{c}$ and $2W^{3}_{c}$.
    }
    \label{layers}
\end{figure}

In this paper we study Kondo effects and  short-range magnetic correlations in a heterostructure at \textit{half-filling} regime, composed by  three interacting square layers, see  Fig. \ref{layers}. \textit{Layer 1} is a Kondo-Heisenberg lattice (KHl) with Heisenberg and Kondo interactions, $J_H$ and $J_K$, respectively. The itinerant part is given by $d$-electrons with hopping term $t_d$. \textit{Layer 2} has itinerant states ($c$-electrons) and $t_c$ hopping term. The coupling between $c$- and $d$-electrons is determined by the hopping term $t_{12}$. In \textit{layer 3} the hopping of $c$-electrons is given by $t_c$ and the coupling between $c$- and $d$-electrons is determined by the hopping term $t_{13}$. The two $c$-electron layers create a \textit{sandwich} with the KHl and $t_{12}$ and $t_{13}$ can
be different. Finally, with the variation of hopping terms and the mixing of the layers, we can study the proximity effects among interfaces in the Kondo state and short-range magnetic correlations of the Kondo-Heisenberg heterostructure in two limiting cases, $|J_{k}|<|J_{H}|$ and $|J_{k}|>|J_{H}|$. 

This article is organized as follow: the theory and model are given in section \ref{Section1}. We find the Green's function that describe the most important self-consistent parameters: $\lambda_{11}$, $\Gamma_{11}$, $\sigma_{12}$ and $\sigma_{13}$ and their phase diagrams in the mean-field approximation. In section \ref{Section2} we show the main numerical results: behaviors of mean-field parameters under proximity effects and the phase diagrams, as well as relations to obtain
the physical quantities of interest. Finally, in section \ref{Section3} we make a critical discussion of the main results. 
%-----------------------------------------------------------------------------------------------
\section{Theory and model}
\label{Section1}

 We aim to describe the proximity effect between three layers where, 
 the first layer, denoted \textit{layer 1}, has itinerant $d$- and localized $f$-electrons coupled through Kondo and Heisenberg interactions. The other two layers (\textit{layer 2} and \textit{layer 3}) are composed of non-interacting itinerant $c$-electrons. We also consider the hopping of electrons between $d$- and $c$-electrons among the three layers. This specific Kondo-Heisenberg heterostructure is shown in Fig. (\ref{layers}). The Hamiltonian that describes this heterostructure is composed of the following three terms
\begin{equation}
    H=H_{\text{KHLL}} + H_{\text{NIL}} +  H_{\text{inter}}.
\label{1}
\end{equation}

The term $H_{\text{KLL}}$
%in Hamiltonian given in Eq.(\ref{1}) 
corresponds to \textit{layer 1} and it is given as
\begin{align}
    H_{\text{KHLL}}=\sum_{\textbf{k}\sigma}  \epsilon_{d1}(\textbf{k})  \hat{n}^{d}_{\textbf{k}1\sigma}+ E_{0} \sum_{i\sigma}  \hat{n}^{f}_{i1\sigma} - \nonumber \\  J_{K}  \sum_{i}   \hat{\textbf{s}}_{d1i}  \cdot \hat{\textbf{S}}_{f1i}- J_{H}\sum_{\langle ij \rangle l} \hat{\textbf{S}}_{f1i} \cdot \hat{\textbf{S}}_{f1j}.
\label{eq3}
\end{align}
Since we are interested in short-range magnetic correlations, the sum over $\langle i,j\rangle$ refers only to nearest-neighbor sites in \textit{layer 1}. $\hat{n}^{d}_{\textbf{k}1\sigma}=d^{\dagger}_{\textbf{k}1\sigma}d_{\textbf{k}1\sigma}$ with $d^{\dagger}_{\textbf{k}1\sigma}$($d_{\textbf{k}1\sigma})$ are creation (annihilation) operators of itinerant $d$-electrons  with \textbf{k}-momentum and spin $\sigma$, while $\hat{n}^{f}_{i1\sigma}=f^{\dagger}_{i1\sigma}f_{i1\sigma}$ with $f^{\dagger}_{i1\sigma}(f_{i1\sigma})$ are creation (annihilation) operators of localized $f$-electrons with spin $\sigma$ at site $i$. $\epsilon^{d}_{1}(\textbf{k})$ is the dispersion relation for $d$-electrons. The third and fourth terms in $ H_{\text{KLL}}$ stand for the Kondo and Heisenberg interactions, respectively, with
 $J_K<0$ and $J_H<0$. The term $H_{\text{NIL}}$, corresponding to two normal layers with non-interacting itinerant $c$-electrons, is  given as
\begin{align}
H_{\text{NIL}}=\sum_{\textbf{k}\sigma}\sum_{l=2}^3\epsilon_{c}(\textbf{k}) \hat{n}^c_{\textbf{k}l\sigma} 
\label{2}
\end{align}
where $\hat{n}^c_{\textbf{k}l\sigma}=c^{\dagger}_{\textbf{k}l\sigma}c_{\textbf{k}l\sigma}$ and $c^{\dagger}_{\textbf{k}l\sigma}$ $(c_{\textbf{k}l\sigma})$ are the creation (annihilation) operator in layer $l=1,2$ with \textbf{k}-momentum 
and spin $\sigma$. We assume a tight-binding conduction band for $d$- and $c$-electrons with bandwidth
$2W_d$ and $2W_c$, respectively. Thus 

\begin{align}    
    \epsilon_{d(c)}(\textbf{k})= -\frac{W_{d(c)}}{2}\sum_{\textbf{R}}\cos(\textbf{k}\cdot\textbf{R})=\nonumber\\- \frac{W_{d(c)}}{2}(\cos(k_x)+\cos(k_y)).
    \label{Eq4}
\end{align}

The coupling between the pairs of layers is given by a tight-binding model, and is expressed as 
\begin{align}
H_{\text{inter}}= -t_{12}\sum_{i j;\sigma}(
d_{i1\sigma}^{\dagger}
c_{j2\sigma} + c^{\dagger}_{i2\sigma} d_{j1\sigma}\nonumber\\ -t_{13}\sum_{i j;\sigma}(
d_{i1\sigma}^{\dagger}
c_{j3\sigma} + c^{\dagger}_{i3\sigma} d_{j1\sigma})
\label{4}
\end{align}
where 
$d^{\dagger}_{i1}$ ($d^{\dagger}_{i1}$)  and  $c^{\dagger}_{i2(3)}$ ($c_{i2(3)}$) are creation (annihilation) operators in site $i$ of \textit{layer 1}, \textit{layer 2} and \textit{layer 3}, respectively. We introduce the hermitian operators to describe the Kondo effect and short-range magnetic correlations \cite{Iglesiass1997, Ruppental1999}, which are given by
\begin{equation*}
    \hat{\lambda}_{i11\sigma}= \frac{1}{2N}(c^{\dagger}_{i1\sigma}f_{i1\sigma}+f^{\dagger}_{i1\sigma} c_{i1\sigma}),
    \label{5}
\end{equation*}

\begin{equation*}
    \hat{\Gamma}_{ij1\sigma}= \frac{1}{2N}(f^{\dagger}_{i1\sigma}f_{j1\sigma}+f^{\dagger}_{j1\sigma} f_{i1\sigma}).
    \label{6}
\end{equation*}

Thus, the spin-flipping part of Kondo and Heisenberg interactions is decoupled as
\begin{equation}
\hat{s}_{d1i}^{x}\hat{S}_{f1i}^{x}+\hat{s}_{d1i}^{y}\hat{S}_{f1i}^{y}=-\sum_{\sigma}\hat{\lambda}_{i11\sigma}\hat{\lambda}_{i11\Bar{\sigma}},
\end{equation}
\begin{equation}
S_{f1i}^{x}S_{f1j}^{x}+S_{f1i}^{y}S_{f1j}^{y}=-\sum_{\sigma}\hat{\Gamma}_{ij1\sigma}\hat{\Gamma}_{ij1\Bar{\sigma}},
\end{equation}
while the $z$ components become 
\begin{equation}
\hat{s}_{d1i}^{z}\hat{S}_{f1i}^{z}=1/4 -\sum_{\sigma}
(\hat{\lambda}_{i1\sigma})^2,
\label{AA61}
\end{equation}
\begin{equation} \hat{S}_{f1i}^{z}\hat{S}_{f1j}^{z}=1/4-\sum_{\sigma}(\hat{\Gamma}_{ij1\sigma})^2.
\label{AA71}
\end{equation}
In Eqs. (\ref{AA61})-(\ref{AA71}), the constraint $\hat{n}^f_{i1}=\hat{n}^{f}_{i1\uparrow}+\hat{n}^{f}_{i1\downarrow}=1$ has been used explicitly. 

We carry out a mean-field treatment of $H_{KLL}$ which neglects quadratic fluctuations $\Delta \hat{\lambda}_{i11\sigma}= (\hat{\lambda}_{i11\sigma}-\lambda_{i11\sigma})$ and $\Delta\hat{\Gamma}_{ij1\sigma}=\hat{\Gamma}_{ij1\sigma}-\Gamma_{ij1\sigma}$  ($\braket{\hat{\lambda}_{i11\sigma}}=\lambda_{i11\sigma}$ and $\braket{\hat{\Gamma}_{ij1\sigma}}=\Gamma_{ij1\sigma}$). Moreover, we consider
$\lambda_{i11\uparrow}=\lambda_{i11\downarrow}=\lambda_{11}$ and similarly $\Gamma_{ij1\uparrow}=\Gamma_{ij1\downarrow}=\Gamma_{11}$, i.e., translational invariant and since there is no magnetic long-range order, also symmetric up and down spin solutions. 
Therefore, for the Heisenberg term, as $\langle i,j\rangle $ refers to  nearest neighbors, we get
\begin{equation*}
    (J_{H}\sum_{\braket{ij}}
  {\hat{\textbf{S}}_{f,i}  \cdot \hat{\textbf{S}}_{f,j})}_{MF}= B \sum_{\textbf{k}\sigma}\epsilon_{d}(\textbf{k})f^{\dagger}_{\textbf{k}\sigma}f_{\textbf{k}\sigma},
\end{equation*}
where
\begin{equation*}
    B= - \frac{4 J_{H}\Gamma}{W^{d}}.
\end{equation*}

As a consequence, the $3$-layer model mean-field Hamiltonian becomes
\begin{widetext}
    
\begin{align}
    (\mathcal{H})_{MF} = & \sum_{\textbf{k},\sigma}(B\epsilon_{d,1}(\textbf{k})+E_{0})f_{\textbf{k}1\sigma}^{\dagger}f_{\textbf{k}1\sigma} + 
\sum_{\textbf{k},\sigma}(\epsilon_{d,1}(\textbf{k}) -\mu)d_{\textbf{k}1\sigma}^{\dagger}d_{\textbf{k}1\sigma}+ \sum_{\textbf{k},\sigma}(\epsilon_{c,2}(\textbf{k})-\mu)c_{\textbf{k}2\sigma}^{\dagger}c_{\textbf{k}2\sigma} \nonumber
\\ 
+ & 2 J_{K}\lambda_{11} \sum_{\textbf{k},\sigma} (d_{\textbf{k}1\sigma}^{\dagger}f_{\textbf{k}1\sigma} + f_{\textbf{k}1\sigma}^{\dagger}d_{\textbf{k}1\sigma}  ) %t^{fc}\sum_{\textbf{k}}(f_{\textbf{k}1\sigma}^{\dagger}c_{\textbf{k}2\sigma} + c_{\textbf{k}2\sigma}^{\dagger}f_{\textbf{k}1\sigma} )
-t_{d1c2}\sum_{\textbf{k},\sigma}(d_{\textbf{k}1\sigma}^{\dagger}c_{\textbf{k}2\sigma} + c_{\textbf{k}2\sigma}^{\dagger}d_{\textbf{k}1\sigma} )-t_{d1c3}\sum_{\textbf{k},\sigma}(d_{\textbf{k}1\sigma}^{\dagger}c_{\textbf{k}3\sigma} + c_{\textbf{k}3\sigma}^{\dagger}d_{\textbf{k}1\sigma} ) \label{HM} \\ 
- & 4J_{k}\lambda_{11}^{2}-2J_{H}\Gamma_{11}^{2}\nonumber.
\end{align}
\end{widetext}

Note that in Eq. (\ref{HM}), the Heisenberg interaction gives rise to a dispersion relation for the $f$-electrons, just as the Kondo interaction leads to effective hybridization between $d$- and $f$-electrons. In both cases, there is a dependence on temperature through $\Gamma_{11}$ giving short-range magnetic correlations and $\lambda_{11}$ that gives the global formation of Kondo singlets which, at high temperatures, can be expected to be null.

The 3-layer mean-field Hamiltonian given in Eq. (\Ref{HM}) is replaced by 
\begin{equation} 
 (\mathcal{H})_{MF}=\sum_{\textbf{k}\sigma} \  \mathbf{\Psi^{\dagger}_{\textbf{k}\sigma}} \ \mathbf{h_{\textbf{k}}} \ \mathbf{\Psi_{\textbf{k}\sigma}} - 4J_{k}\lambda_{11}^{2}-2J_{H}\Gamma_{11}^{2}, 
\label{A0}
\end{equation}
where

\[
\mathbf{\Psi_{\textbf{k}\sigma}}=\left( \begin{array}{c}
 f_{\textbf{k}1\sigma} \\ 
 d_{\textbf{k}1\sigma} \\ 
 c_{\textbf{k}2\sigma}\\ 
  c_{\textbf{k}3\sigma}\\
\end{array} \right)  
\]
and 
\[
\mathbf{h_\textbf{k}}= 
\left( \begin{array}{cccc}
B\epsilon_{d1}(\textbf{k}) +E_{0}& 2J_{k}\lambda_{11}             & 0                     & 0 \\ 
2J_{k}\lambda_{11}                & \epsilon_{d1}(\textbf{k}) & -t_{dc_{2}}           & -t_{dc_{3}}\\
0                                 & -t_{dc_{2}}               & \epsilon_{c,2}(\textbf{k}) & 0\\
0                                 & -t_{dc_{3}}         & 0          & \epsilon_{c,2}(\textbf{k})
\end{array} 
\right). 
\]
%

%%%%%%%%%%

Using the results obtained in Eqs. (\Ref{A7})-(\Ref{D3}), see Appendix \ref{appendix}, the %correlation functions 
correlator $\lambda_{11}$, $\Gamma_{11}$,  $\sigma_{12}$ and $\sigma_{13}$ can be obtained self-consistently from the coupled equations
\begin{equation}
\lambda_{11} =  \frac{1}{N} \sum_{\textbf{k}}
 \oint \frac{dz}{2\pi i} f(z)\braket{\braket{ d_{\textbf{k},1\sigma} :f^{\dagger}_{\textbf{k},1,\sigma }  }}_{z},
\label{L11}
\end{equation}
\begin{equation}
1 = \frac{J_H}{W^d} \frac{1}{N} \sum_{\textbf{k}}
\epsilon_d(\textbf{k}) \oint \frac{dz}{2\pi i} f(z)\braket{\braket{ f_{\textbf{k},1\sigma} :f^{\dagger}_{\textbf{k},1,\sigma }  }}_{z}
\label{G}
\end{equation}
and
\begin{equation}
\sigma_{12} =  \frac{1}{N} \sum_{\textbf{k}}
\oint \frac{dz}{2\pi i} f(z)\braket{\braket{ c_{\textbf{k},2\sigma} :d^{\dagger}_{\textbf{k},1,\sigma }  }}_{z},
\label{L12}
\end{equation}
\begin{equation}
\sigma_{13} =  \frac{1}{N} \sum_{\textbf{k}}
\oint \frac{dz}{2\pi i} f(z)\braket{\braket{ c_{\textbf{k},3\sigma} :d^{\dagger}_{\textbf{k},1,\sigma }  }}_{z}.
\label{L12}
\end{equation}
 
 The contour of the path integral in Eqs. (\Ref{L11}) - (\Ref{L12}) encircles the real axis without enclosing any poles of the Fermi-Dirac distribution.
We also fix the total electronic density $n_{t}=(\langle \hat{n}_d \rangle +\langle \hat{n}_{c2}\rangle + \langle \hat{n}_{c3}\rangle )/3$ and the \textit{f}-level $E_0$ using the chemical potential $\mu$. To complete the self-consistency,  
%and replacing 
the constraint $\hat{n}_f=1$ is replaced by the mean field one $\langle \hat{n}_f \rangle =1$.
%, respectively.

\section{Results and discussion}\label{Section2}

To simplify the complexity of numerical results, we consider the bandwidth given by: $W^{d}=W^{c}_{2(3)}=1$.  The density of states for both the  $d$- and $c$-electrons is assumed to be semi-parabolic, and is given by
\begin{equation}
D(\omega)= \frac{3}{4}(1-\epsilon^{2}) \ \ |\epsilon|\leq 1.
\label{DOS}
\end{equation}
The energies are measured in
units of the half bandwidth  to obtain dimensionless parameters and subsequently normalized by $t_{c}/W^{c}_{2(3)}$. In the following, to study the correlation effects in the heterostructure, we display the self-consistent solutions for the mean-field parameters given in Eqs. (\ref{L11})-(\ref{L12}) versus the interlayer hopping parameters $t_{12}/t_{c}$ and $t_{13}/t_{c}$ at \textit{half-filling} regime, i.e., the occupation number of $f$-electrons in the KHl is $n^f=1$, where $n^f=f^{\dagger}_{\textbf{k}1\sigma}f_{\textbf{k}1\sigma}$, $t_{c}=1$ and $k_{B}=1$.

We are interested in two limiting cases, $|J_{K}|<|J_{H}|$ and $|J_{K}|>|J_{H}|$. In the case $|J_{H}|>|J_{K}|$, the non-local coupling between $f$-electrons is much stronger than the non-local coupling between $d$- and $f$-electrons.  In the case $|J_{K}|>|J_{H}|$, the non-local interactions between $d$- and $f$-electrons dominate over the non-local coupling  of $d$-electrons, see Fig. \ref{layers}. The relative size mean-field parameters $\Gamma_{11}/t_{c}$ and $\lambda_{11}/t_{c}$ indicate the relative the stability of the Kondo and short-range magnetic correlations. The quantities $\sigma_{12}/t_{c}$ and $\sigma_{13}/t_{c}$ describe the coupling between the layers conduction bands.

\subsection{Hopping dependence: $t_{12}=t_{13}$}
\begin{figure}
    \centering    \includegraphics[scale=0.38]{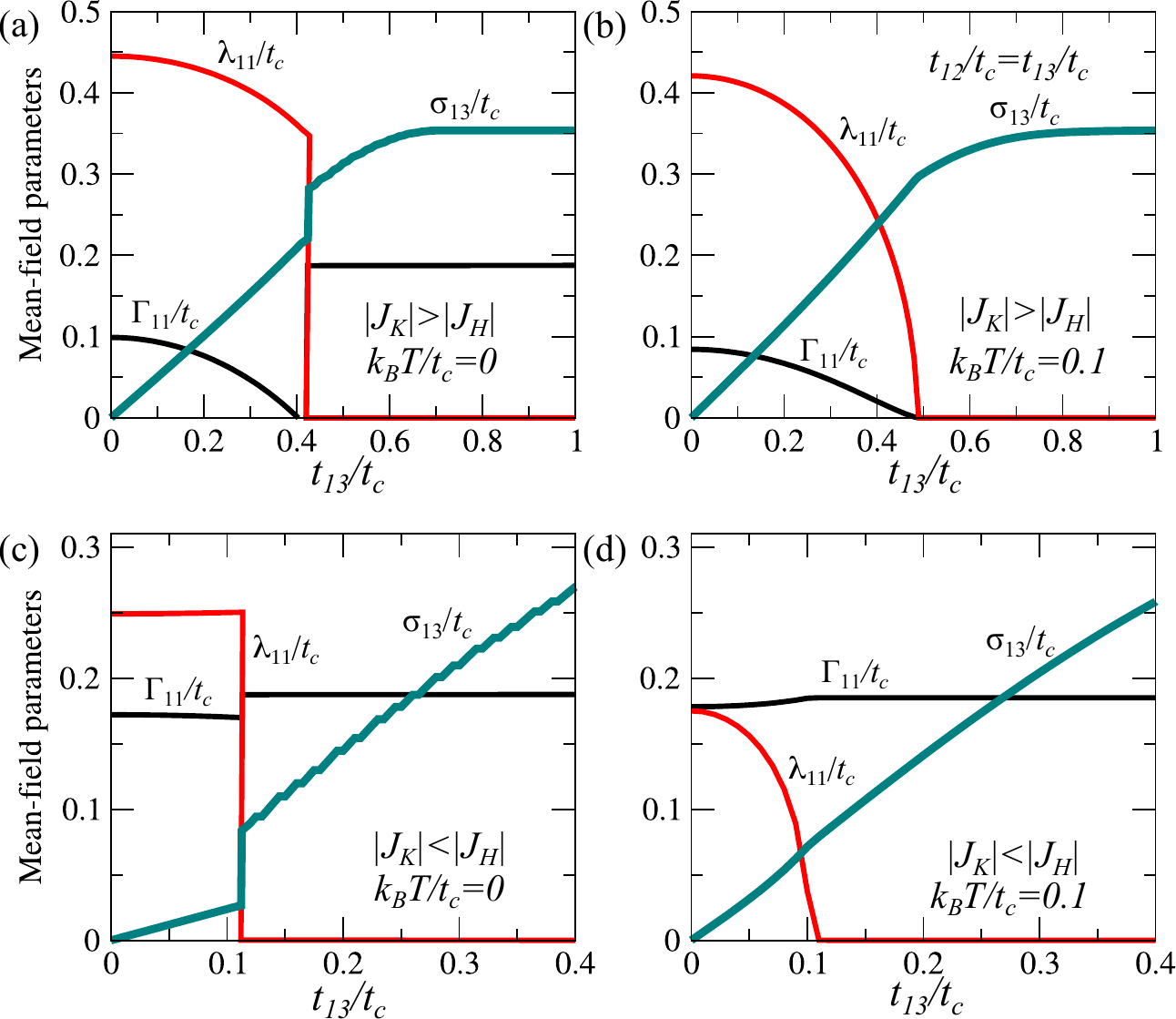}
    \caption{Mean-field parameters $\lambda_{11}/t_{c}$, $\Gamma_{11}/t_{c}$ and $\sigma_{13}/t_{c}$ as functions of hopping parameter $t_{13}/t_{c}$, at the \textit{half-filling} regime. (a)$J_{K}=-0.5$, and $J_{H}=-0.1$ at $k_{B}T/t_{c}=0$, (b) $J_{K}=-0.5$, and $J_{H}=-0.1$ at $k_{B}T/t_{c}=0.1$, (c) $J_{K}=-0.5$, and $J_{H}=-1$ at $k_{B}T/t_{c}=0$ and (d) $J_{K}=-0.5$, and $J_{H}=-1$ at $k_{B}T/t_{c}=0.1$. $W^{c}_{2(3)}=W^{d}=1$. The hopping terms are equal to $t_{12}/t_{c}=t_{13}/t_{c}$. The behavior of $\sigma_{12}$ is similar to $\sigma_{13}$.  }
    \label{Rup0}
\end{figure}

In Fig. \ref{Rup0}, we show the mean-field parameters $\Gamma_{11}/t_{c}$, $\lambda_{11}/t_{c}$, $\sigma_{12}/t_{c}$ and $\sigma_{13}/t_{c}$ as functions of hopping term $t_{12}/t_{c}=t_{13}/t_{c}$. In Figs. \ref{Rup0}(a) and \ref{Rup0}(c), the mean-field parameters, at $k_{B}T/t_{c}=0$, are found for $|J_{K}|< |J_{H}|$ and $|J_{K}|> |J_{H}|$.  In Figs. \ref{Rup0}(a) and \ref{Rup0}(c) the parameter $\lambda_{11}/t_{c}$ decreases sharply at $t_{13}/t_{c}\approx 0.4$ and $t_{13}/t_{c}\approx 0.1$, respectively, disappearing for higher magnitudes of $t_{13}/t_{c}$. When $|J_{K}|<
|J_{H}|$, the parameter $\Gamma_{11}/t_{c}$ continuously decreases until reaching zero for $t_{13}/t_{c}\approx 0.4$ and remains constant for $t_{13}/t_{c}>0.4$. In contrast, when $|J_{K}|> |J_{H}|$, $\Gamma_{11}/t_{c}$ remains constant for all $t_{13}/t_{c}$ but undergoes a sudden change at $t_{13}/t_{c}\approx 0.1$. The parameters $\sigma_{12}/t_{c}$ and $\sigma_{13}/t_{c}$ are equal and increase linearly but undergo an abrupt discontinuity at $t_{13}/t_{c}\approx 0.4$ and $t_{13}/t_{c}\approx 0.1$, continue to show a linear increase with $t_{13}/t_{c}$. In Figs. \ref{Rup0}(b) and \ref{Rup0}(d), the mean-field parameters at $k_{B}T/t_{c}=0.1$ are shown for $|J_{K}|< |J_{H}|$ and $|J_{K}|> |J_{H}|$, respectively. When $|J_{K}|< |J_{H}|$, the parameters $\Gamma_{11}/t_{c}$ and $\lambda_{11}/t_{c}$ decreases quadratically at $t_{13}/t_{c}\approx 0.4$, while the parameters $\sigma_{12}/t_{c}$ and $\sigma_{13}/t_{c}$ are equal, i.e., $\sigma_{12}/t_{c}=\sigma_{13}/t_{c}$, increasing linearly until stabilizing and becoming constants for high values of $t_{13}/t_{c}$.  When $|J_{K}|> |J_{H}|$, the parameter $\lambda_{11}/t_{c}$ continuously decreases at $t_{13}/t_{c}\approx 0.1$, the parameter $\Gamma_{11}/t_{c}$ is constant and independent of $t_{13}/t_{c}$, while $\sigma_{12}/t_{c}=\sigma_{13}/t_{c}$ are increasing linearly with the increase of $t_{13}/t_{c}$.

%%%%%%%%%%%%%%%%%%%%%%%%%%%%%%%%%%%%%%%%%%%%%%%%%%%%%%%%%%%%%%%%%%%%%%%%%%%%%%%%%%%%%%%

\subsection{Hopping dependence: $t_{12}\not=t_{13}$}
\begin{figure}
    \centering    \includegraphics[scale=0.39]{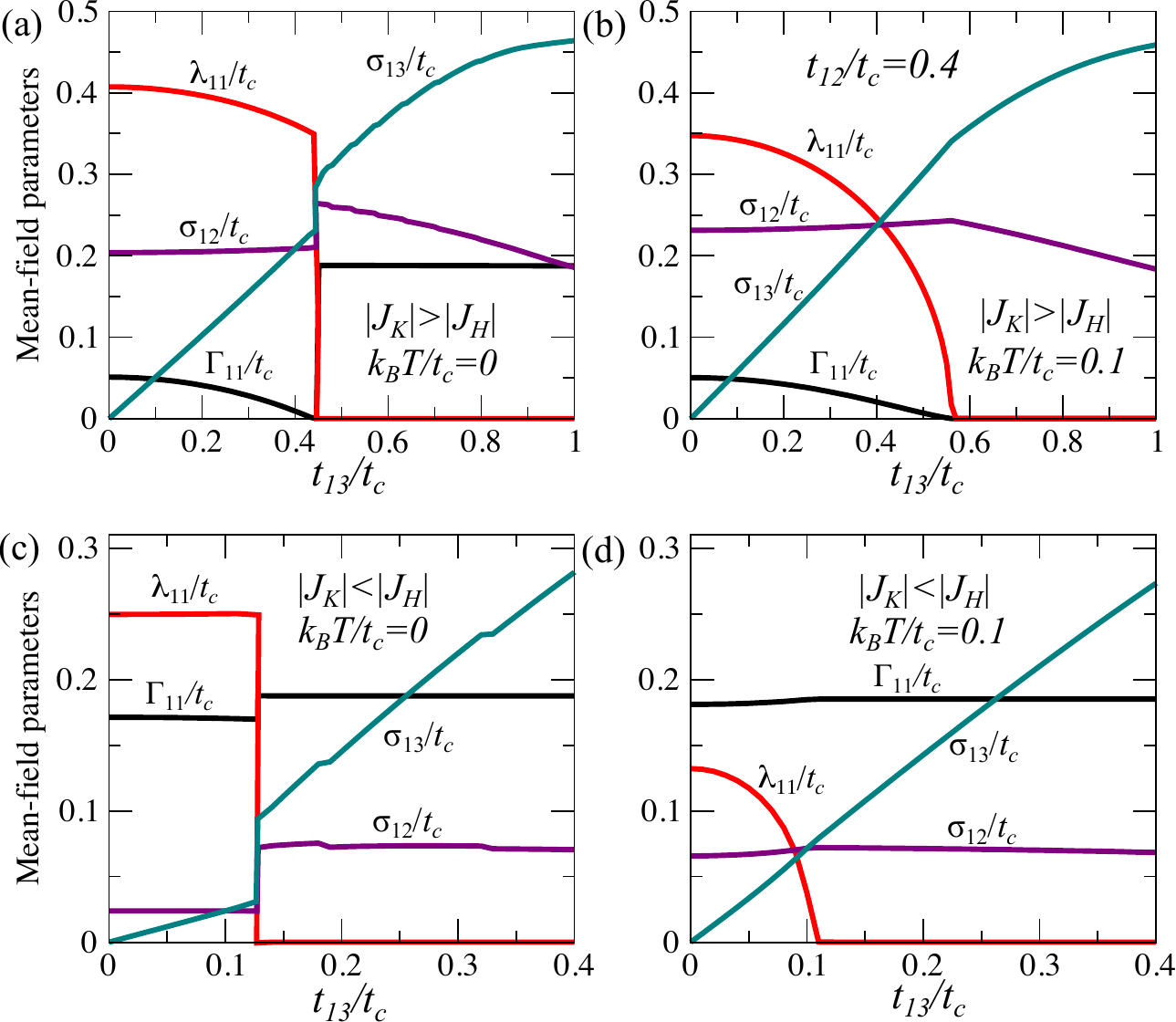}
    \caption{Mean-field parameters $\lambda_{11}/t_{c}$, $\Gamma_{11}/t_{c}$, $\sigma_{12}/t_{c}$ and $\sigma_{13}/t_{c}$ as functions of hopping term $t_{13}/t_{c}$, for $n^{f}=1$. (a) $J_{K}=-0.5$, and $J_{H}=-0.1$ at $k_{B}T/t_{c}=0$, (b) $J_{K}=-0.5$, and $J_{H}=-0.1$ at $k_{B}T/t_{c}=0.1$, (c) $J_{K}=-0.5$, and $J_{H}=-1$ at $k_{B}T/t_{c}=0$ and (d)$J_{K}=-0.5$, and $J_{H}=-1$ at $k_{B}T/t_{c}=0.1$. $W^{c}_{2(3)}=W^{d}=1$. The hopping term $t_{12}/t_{c}=0.4$.}
    \label{Rup1}
\end{figure}

Fig. \ref{Rup1} shows the mean-field parameters $\Gamma_{11}/t_{c}$, $\lambda_{11}/t_{c}$, $\sigma_{12}/t_{c}$, and $\sigma_{13}/t_{c}$ as functions of the hopping term $t_{13}/t_{c}$, at $k_{B}T/t_{c}=0$ and $k_{B}T/t_{c}=0.1$.  The hopping terms $t_{12}/t_{c}$ and $t_{13}/t_{c}$ differ, allowing the distances between \textit{layer 1} and \textit{layer 2} to be distinct from \textit{layer 1} and \textit{layer 3}. For simplicity, we had chosen $t_{12}/t_{c}=0.4$. The behavior of the order parameters $\Gamma_{11}/t_{c}$ and $\lambda_{11}/t_{c}$ are essentially the same as shown in Fig. \ref{Rup0}, but with different critical values of $t_{13}/t_{c}$. For $k_{B}T/t_{c}=0$, shown in Figs. \ref{Rup1}(a) and \ref{Rup0}(c), the parameters $\sigma_{12}/t_{c}$ and $\sigma_{13}/t_{c}$ exhibit different behaviors. $\sigma_{12}/t_{c}$ remains constant until $t_{13}/t_{c}\approx0.4$, undergoes a small abrupt variation and subsequently decays linearly with increasing $t_{13}/t_{c}$. $\sigma_{13}/t_{c}$ linearly increases, experiencing a discontinuity at $t_{13}/t_{c}\approx0.4$, and subsequently continues to grow.  When $|J_{K}|<|J_{H}|$, the behavior of $\sigma_{12}/t_{c}$ is constant and independent of $t_{13}/t_{c}$, but experiences a slight discontinuity around $t_{13}/t_{c}\approx0.1$.  At $k_{B}T/t_{c}=0.1$, as depicted in Figs. \ref{Rup1}(b) and \ref{Rup1}(d), when $|J_{K}|>|J_{H}|$, the parameter $\sigma_{12}/t_{c}$ remains nearly constant until $t_{13}/t_{c}\approx0.6$ and then decays linearly as $t_{13}/t_{c}$ increases. The parameter $\sigma_{13}/t_{c}$ increases for all values of $t_{13}/t_{c}$.  When $|J_{K}|<|J_{H}|$, the parameter $\sigma_{12}/t_{c}$ is nearly constant and independent of $t_{13}/t_{c}$, the parameter $\sigma_{13}/t_{c}$ increase continuously for all values of $t_{13}/t_{c}$. In the following we show numerical results regarding the behavior of the mean-field parameters $\Gamma_{11}/t_{c}$ and $\lambda_{11}/t_{c}$ with the variation of the interlayer hopping terms.

\subsection{ Temperature dependence: case $|J_{K}|<|J_{H}|$}

\begin{figure}
    \centering    \includegraphics[scale=0.39]{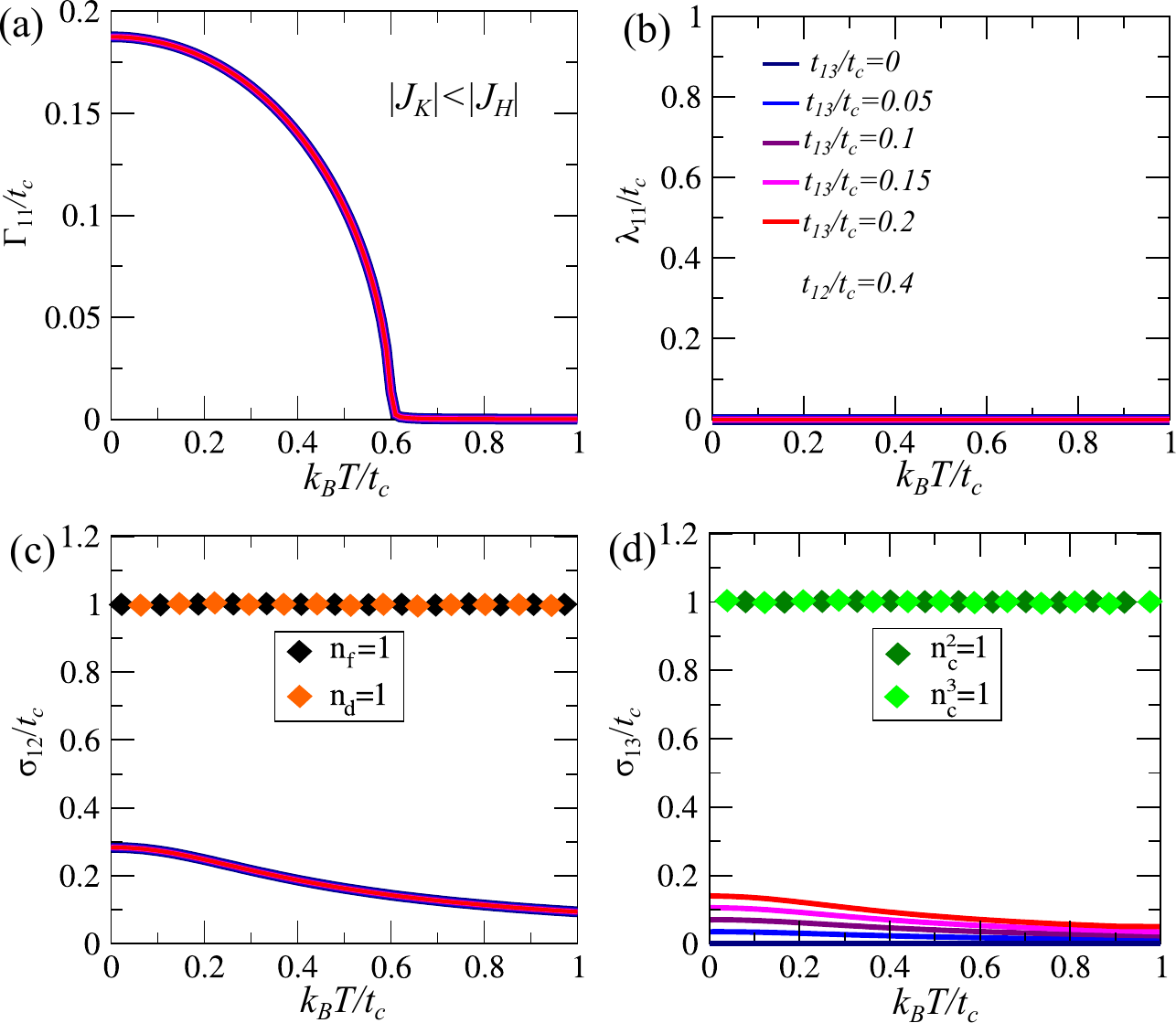}
   \caption{Mean-field parameters (a) $\lambda_{11}/t_{c}$,  (b) $\Gamma_{11}/t_{c}$, (c) $\sigma_{12}/t_{c}$ and (d) $\sigma_{13}/t_{c}$ as function of $k_{B}T/t_{c}$ under a variation of $t_{13}/t_{c}$ for $n^{f}=1$ and $n_{t}=(n_{d}+n^{2}_{c}+n^{3}_{c})/3=1$ in the case where the
Heisenberg interaction is large compared to the Kondo coupling:
$J_{K}=-0.5$, and $J_{H}=-1$. $W^{c}_{2(3)}=W^{d}=1$ and the hopping parameter $t_{12}/t_{c}=0.4$. }
    \label{Rup5}
\end{figure}

%%%%%%%%%%%%%%%%%%%%%%%%%%%%%%%%%%%%%%%%%%%%%%%%%%%%%%%%%%%%%%%%%%%%%%%%%%%%%%%%%%%%%%%%%%%%%%
Fig. \ref{Rup5} displays the mean-field parameters $\Gamma_{11}/t_{c}$, $\lambda_{11}/t_{c}$, $\sigma_{12}/t_{c}$, and $\sigma_{13}/t_{c}$ as functions of $k_{B}T/t_{c}$, with varying hopping $t_{13}$.  The hopping term between \textit{layer 1} and \textit{layer 2} is fixed at $t_{12}/t_{c}=0.4$, maintaining a constant distance between these layers, while the distance between \textit{layer 1} and \textit{layer 3} varies, allowing for the finite variation of $t_{13}/t_{c}$. In Fig. \ref{Rup5}(a), the behavior of the mean-field parameter $\Gamma_{11}/t_{c}$ is shown as a function of $k_{B}T/t_{c}$ for finite values of $t_{13}/t_{c}$ and $t_{12}/t_{c}=0.4$. We can find that the behavior of $\Gamma_{11}/t_{c}$ is independent of the values of $t_{13}/t_{c}$ and quadratically decays around $k_{B}T/t_{c} \approx 0.6$. The mean-field parameter $\lambda_{11}/t_{c}$ is zero in this case and is independent of $t_{13}/t_{c}$ (see Fig. \ref{Rup5}(b)). Figs. \ref{Rup5}(c) and \ref{Rup5}(d) show the parameters $\sigma_{12}/t_{c}$ and $\sigma_{13}/t_{c}$, respectively, along with the occupation numbers of the $f$-, $d$-, and $c$-electrons in each of the three layers.  The parameter $\sigma_{12}/t_{c}$ asymptotically decays with increasing $k_{B}T/t_{c}$ and remains constant with respect to variations of $t_{13}/t_{c}$.  The parameter $\sigma_{13}/t_{c}$ also asymptotically decays with increasing $k_{B}T/t_{c}$, with a slight variations on changing $t_{13}/t_{c}$. 

\begin{figure}
    \centering    \includegraphics[scale=0.39]{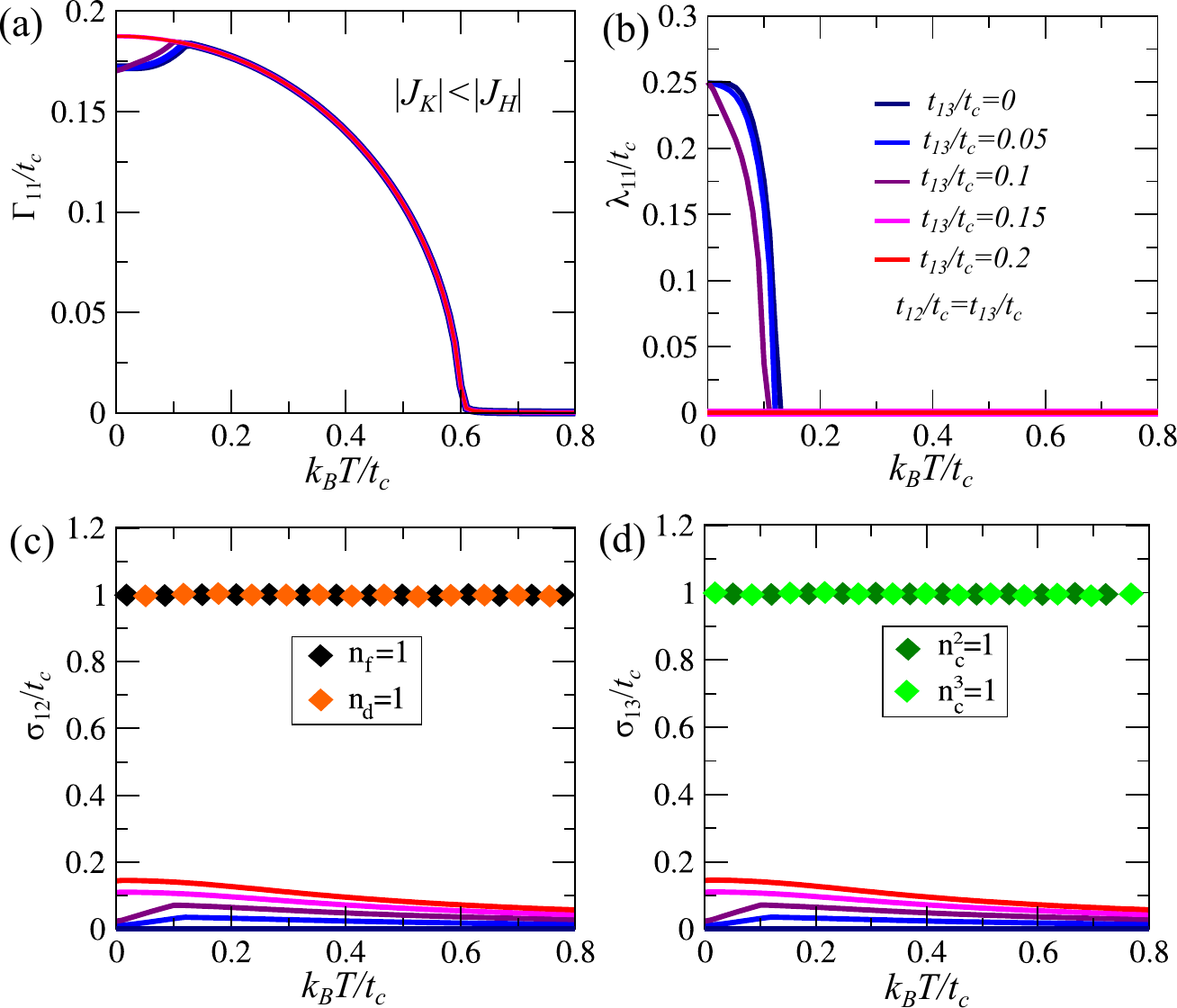}
   \caption{Mean-field parameters (a) $\lambda_{11}/t_{c}$,  (b) $\Gamma_{11}/t_{c}$, (c) $\sigma_{12}/t_{c}$ and (d) $\sigma_{13}/t_{c}$ as function of $k_{B}T$ under a variation of $t_{13}/t_{c}$ for $n^{f}=1$ and $n=(n_{d}+n^{2}_{c}+n^{3}_{c})/3=1$ in the case where the
Heisenberg interaction is large compared to the Kondo coupling:
$J_{K}=-0.5$, and $J_{H}=-1$. $W^{c}_{2(3)}=W^{d}=1$ and the hopping parameters are $t_{12}/t_{c}=t_{13}/t_{c}$. }
    \label{Rup3}
\end{figure}

Fig. \ref{Rup3} shows the mean-field parameters $\Gamma_{11}/t_{c}$, $\lambda_{11}/t_{c}$, $\sigma_{12}/t_{c}$, and $\sigma_{13}/t_{c}$ as functions of $k_{B}T/t_{c}$ under a finite variation of $t_{12}/t_{c}=t_{13}/t_{c}$ at $n^{f}=1$ and $n_{t}=(n_{d}+n^{2}{c}+n^{3}{c})/3=1$. Similar to the previous result, Fig. \ref{Rup5}(a), the self-consistent mean-field parameter $\Gamma_{11}/t_{c}$, Fig. \ref{Rup3}(a), continuously decreases for all magnitudes of $t_{12}/t_{c}$, reaching vanishing small values  at $k_{B}T/t_{c}=0.6$, except for values below $k_{B}T/t_{c}\approx 0.1$ where a slight increment is observed for $t_{12}/t_{c}<0.15$. In Fig. \ref{Rup3}(b), the mean-field parameter $\lambda_{11}/t_{c}$ steadily decreases until completely disappearing around $k_{B}T/t_{c}\approx0.1$ for magnitudes lower than $t_{13}/t_{c}<0.15$, while for $t_{13}/t_{c}=0.2$, the mean-field parameter $\lambda_{11}/t_{c}$ is zero. In Figs. \ref{Rup3}(c) and \ref{Rup3}(d), the parameters $\sigma_{12}/t_{c}$ and $\sigma_{13}/t_{c}$, with $\sigma_{12}/t_{c}=\sigma_{13}/t_{c}$, asymptotically decay for all values of $k_{B}T/t_{c}$ and $t_{13}/t_{c}$, while the occupation values of all electrons remain constant and equal to 1, i.e., $n^{f}=1$.

\subsection{Phase diagrams: case $|J_{K}|<|J_{H}|$}

\begin{figure}
    \centering    \includegraphics[scale=0.45]{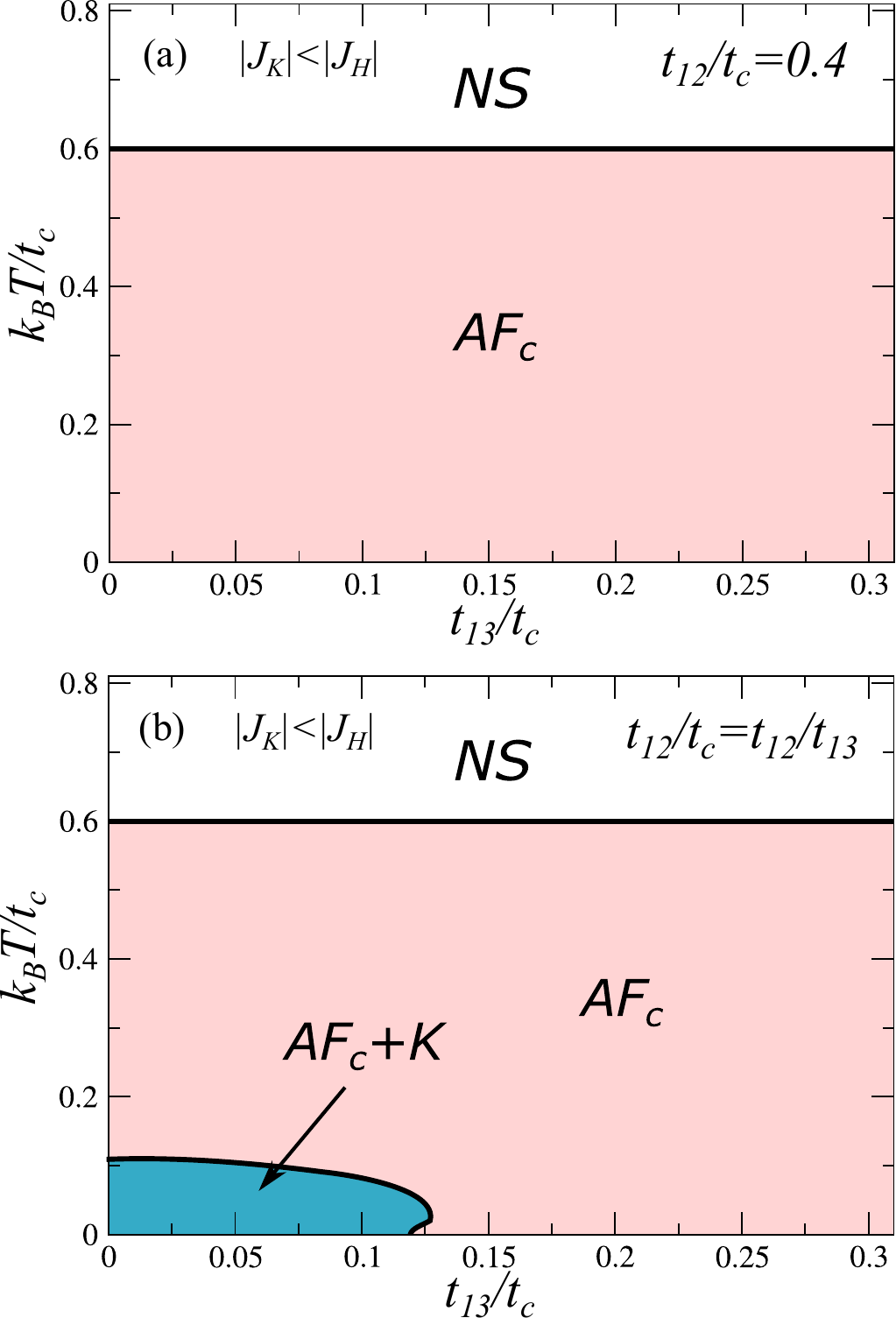}
   \caption{Phase diagrams of $k_{B}T/t_{c}$ versus $t_{13}/t_{c}$ with (a) $t_{12}/t_{c}=0.4$  and (b) $t_{12}/t_{c}=t_{13}/t_{c}$ in the case where the
Heisenberg interaction is large compared to the Kondo coupling:
$J_{K}=-0.5$, and $J_{H}=-0.1$. $W^{c}_{2(3)}=W^{d}=1$. $NS$ is a normal state, $F_{c}$ are ferromagnetic correlations and $K$ is a Kondo state. }
    \label{DF1}
\end{figure}

Although the mean-field parameters are not order parameters since they describe short-ranged correlations, they exhibit behavior akin to order parameters. Based on these parameters, we construct a phase diagram that does not indicate thermodynamic phase boundaries but, instead reflect the different correlations in the phases.
The phase diagrams are constructed from Eqs. (\ref{L11}) and (\ref{G}) for the mean-field parameters $\Gamma_{11}/t_{c}$ and $\lambda_{11}/t_{c}$ in the case $|J_{K}|<|J_{H}|$. In Fig. \ref{DF1}(a), we depict the phase diagram of $k_{B}T/t_{c}$ versus $t_{13}/t_{c}$, for a
fixed value of $t_{12}/t_{c}=0.4$, i.e.,  the interfaces of \textit{layer 1} and \textit{layer 3} are at a fixed distance. In this case, only short-range antiferromagnetic correlations $AF_{c}$ are observed, independent of the value of $t_{13}/t_{c}$. For low values of $t_{13}/t_{c}$ and as $k_{B}T/t_{c}$ decreases, a continuous transition (solid line) from a normal $NS$ to $AF_{c}$ is found, and as $t_{13}/t_{c}$ starts to increase, for high values of $k_{B}T/t_{c}$, a continuous transition from $AF_{c}$ to a $NS$ is observed again. The absence of Kondo correlations is simply due to $J_{K}$ being much lower than $J_{H}$. In addition, in Fig. \ref{DF1}(b), the phase diagram is shown when $t_{12}/t_{c}=t_{13}/t_{c}$. For low values of $t_{13}/t_{c}$ and as $k_{B}T/t_{c}$ decreases, there is a continuous transition from  $NS$ to $AF_{c}$ correlations. The "dome" corresponds to the coexistence of Kondo and $AF_{c}$. The "dome" shows that the Kondo correlations disappear continuously near $t_{13}/t_{c}=0.15$. 
For $k_{B}T/t_{c}$ higher than $t_{13}/t_{c}>0.13$, only  $AF_{c}$ correlations exist, regardless of the value of $t_{13}/t_{c}$.

%%%%%%%%%%%%%%%%%%%%%%%%%%%%%%%%%%%%%%%%%%%%%%%%%%%%%%%%%%%%%%%%%%%%%%%%%%%%%%%%%%%%%%%%%%%%%
 
\subsection{Temperature dependence: case $|J_{K}|>|J_{H}|$}

%%%%%%%%%%%%%%%%%%%%%%%%%%%%%%%%%%%%%%%%%%%%%%%%%%%%%%%%%%%%%%%%%%%%%%%%%%%%%%%%%%%%%%%%

\begin{figure}
    \centering    \includegraphics[scale=0.39]{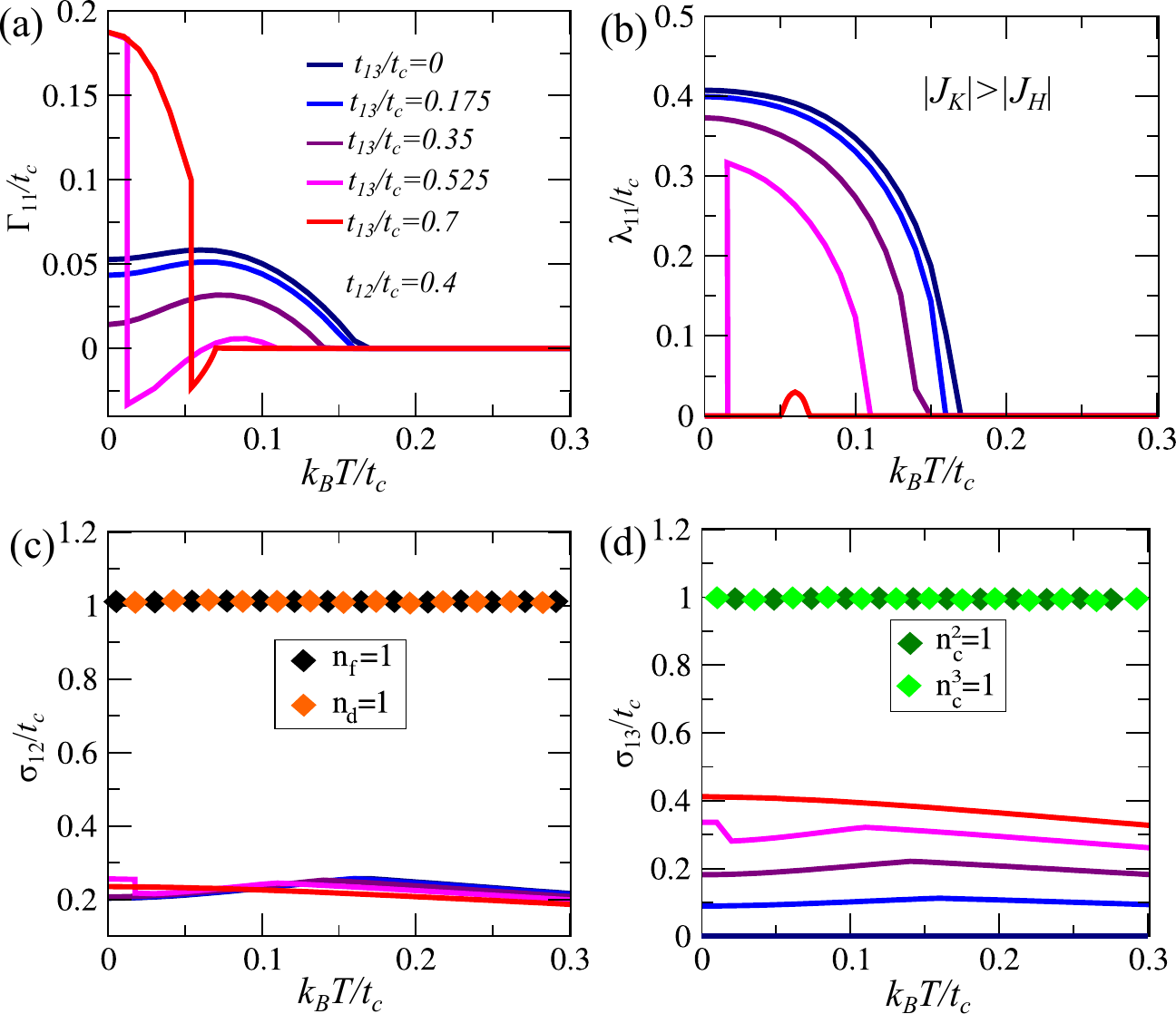}
   \caption{Mean-field parameters (a) $\Gamma_{11}/t_{c}$,  (b) $\lambda_{11}/t_{c}$, (c) $\sigma_{12}/t_{c}$ and (d) $\sigma_{13}/t_{c}$ as function of $k_{B}T/t_{c}$ under a variation of $t_{13}/t_{c}$ for $n^{f}=1$ and $n_{t}=(n_{d}+n^{2}_{c}+n^{3}_{c})/3=1$ in the case where the
Heisenberg interaction is large compared to the Kondo coupling:
$J_{K}=-0.5$, and $J_{H}=-0.1$. $W^{c}_{2(3)}=W^{d}=1$ and the hopping parameter $t_{12}/t_{c}=0.4$. }
    \label{Rup4}
\end{figure}

%%%%%%%%%%%%%%%%%%%%%%%%%%%%%%%%%%%%%%%%%%%%%%%%%%%%%%%%%%%%%%%%%%%%%%%%%%%%%%%%%%%%%%%%%%

Fig. \ref{Rup4} depicts the mean-field parameters $\Gamma_{11}/t_{c}$, $\lambda_{11}/t_{c}$, $\sigma_{12}/t_{c}$ and $\sigma_{13}/t_{c}$ as function of $k_{B}T/t_{c}$ under varying hopping parameter $t_{13}/t_{c}$, for $n^{f}=1$ and $n_{t}=(n_{d}+n^{2}{c}+n^{3}{c})/3=1$. In the case $|J_{K}|>|J_{H}|$, the hopping term between the \textit{layer 1} and \textit{layer 2} is fixed at $t_{12}/t_{c}=0.4$, maintaining a constant distance between these layers, while the distance between the \textit{layer 1} and \textit{layer 2} varies, allowing for the finite variation of $t_{13}/t_{c}$. In Fig. \ref{Rup4}(a), the behavior of the mean-field parameter $\Gamma_{11}/t_{c}$, with respect to $k_{B}T/t_{c}$ is shown under the variation of $t_{13}/t_{c}$. A discontinuity in $\Gamma_{11}/t_{c}$ is observed for estimated values of $0.525<t_{13}/t_{c}<0.7$ at $k_{B}T/t_{c}\approx0.01$ and $k_{B}T/t_{c}\approx0.05$. The negative signal in $\Gamma_{11}/t_{c}$ indicates that the spin correlation between the neighboring sites became ferromagnetic and when $\Gamma_{11}/t_{c}$ goes to zero the spin correlations at different sites will become zero, since the spins are decoupled. The next effect is similar to varying from an antiferromagnetic state to a ferromagnetic state \cite{Ruppental1999}. Furthermore, the mean-field parameter $\lambda_{11}/t_{c}$ as a function of $k_{B}T/t_{c}$ is depicted in Fig. \ref{Rup4}(b). For $t_{13}/t_{c}<0.525$, $\lambda_{11}/t_{c}$ continuously decreases until it reduces to zero. Conversely, for $t_{13}/t_{c}=0.525$, $\lambda_{11}/t_{c}$ is absent for low $k_{B}T/t_{c}$ values and abruptly increases around $k_{B}T/t_{c}\approx0.01$. Specifically, for $t_{13}/t_{c}=0.7$, a small region, or "dome" where $\lambda_{11}/t_{c}$ exist is shown. Figs. \ref{Rup4}(c) and \ref{Rup4}(d) display the parameters $\sigma_{12}/t_{c}$, $\sigma_{13}/t_{c}$, and the respective occupation numbers of the $f$-, $d$-, and $c$-electrons in each of the three layers. We note that $\sigma_{12}/t_{c}\neq\sigma_{13}/t_{c}$ due to the constant distance between \textit{layer 1} and \textit{layer 2}, while the distance between \textit{layer 1} and \textit{layer 3} varies. Despite the differences in magnitude, both parameters asymptotically decay for high $k_{B}T/t_{c}$ values. 

\begin{figure}
    \centering    \includegraphics[scale=0.39]{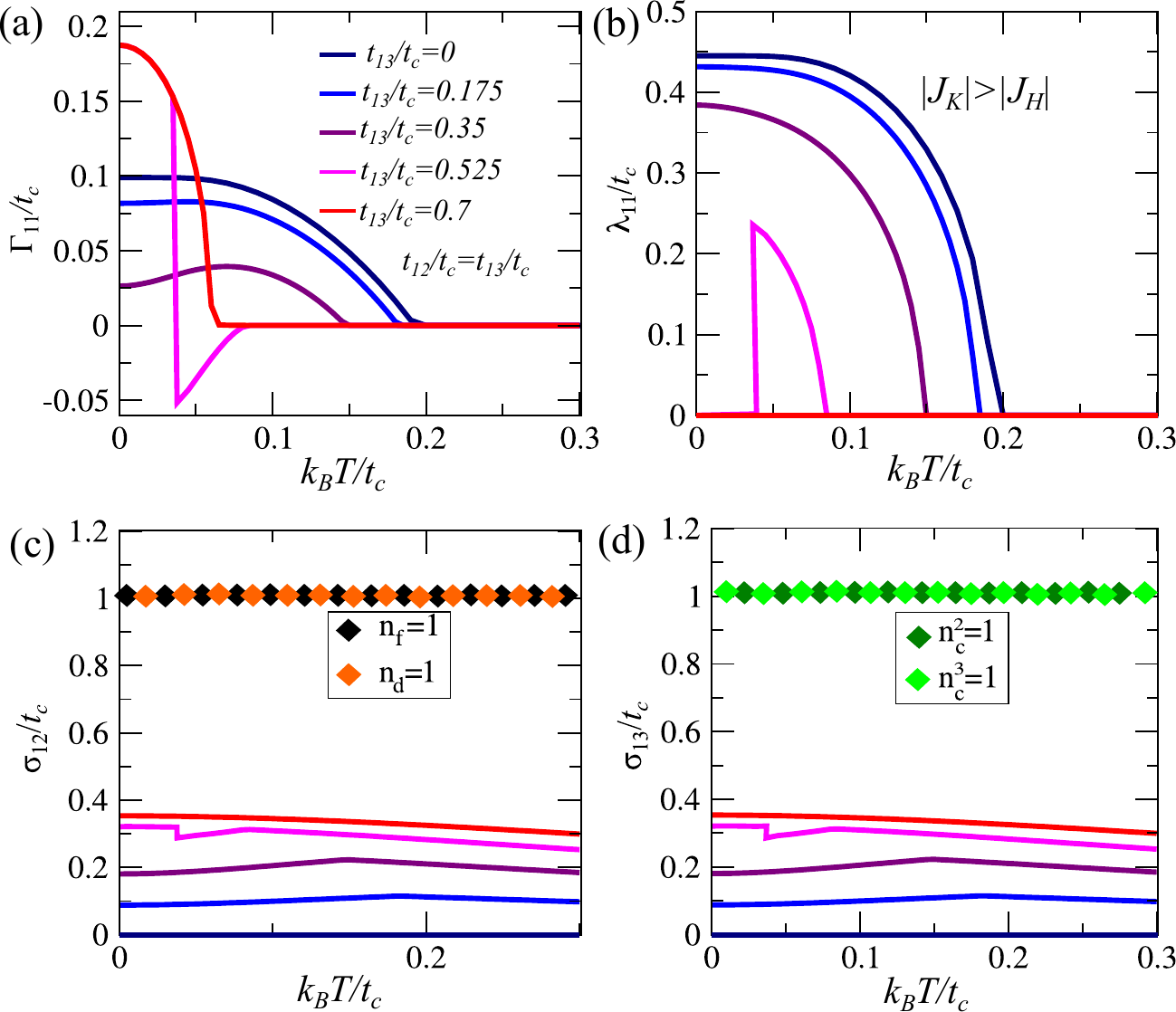}
   \caption{Mean-field parameters (a) $\Gamma_{11}/t_{c}$,  (b) $\lambda_{11}/t_{c}$, (c) $\sigma_{12}/t_{c}$ and (d) $\sigma_{13}$ as function of $k_{B}T/t_{c}$ under a variation of $t_{13}/t_{c}$ for $n^{f}=1$ and $n_{t}=(n_{d}+n^{2}_{c}+n^{3}_{c})/3=1$ in the case where the
Heisenberg interaction is large compared to the Kondo coupling:
$J_{K}=-0.5$, and $J_{H}=-0.1$. $W^{c}_{2(3)}=W_{d}=1$ and the hopping parameters are $t_{12}/t_{c}=t_{13}/t_{c}$. }
    \label{Rup2}    
\end{figure}

The self-consistent mean-field parameters  $\Gamma_{11}/t_{c}$, $\lambda_{11}/t_{c}$, $\sigma_{12}/t_{c}$, and $\sigma_{13}/t_{c}$ as functions of $k_{B}T/t_{c}$ for finite hopping term values, $t_{12}/t_{c}=t_{13}/t_{c}$ are depicted in Fig. \ref{Rup2} for the case $J_{K}|>|J_{H}|$. The parameter $\Gamma_{11}/t_{c}$, Fig. \ref{Rup2}(a), diminishes at lower $k_{B}T/t_{c}$ values. Specifically, at $t_{13}/t_{c}=0.525$, a strong discontinuity occurs at a $k_{B}T/t_{c}\approx0.045$. The parameter $\lambda_{11}/t_{c}$, depicted in Fig. \ref{Rup2}(b), as the value of $t_{13}/t_{c}$ increases, diminishes at lower $k_{B}T/t_{c}$ values. Specifically, at $t_{13}/t_{c}=0.525$ (the same value of $t_{13}/t_{c}$ that causes the discontinuity in $\Gamma_{11}/t_{c}$), the parameter $\lambda_{11}/t_{c}$ only exists between $0.045<k_{B}T/t_{c}<0.085$. When $t_{13}/t_{c}=0.7$, the parameter $\lambda_{11}/t_{c}$ is completely suppressed. In Figs. \ref{Rup2}(c) and \ref{Rup2}(d), the behavior of the parameters $\sigma_{12}/t_{c}=\sigma_{13}/t_{c}$ and the respective layer occupation numbers are shown. It is evident that for different $k_{B}T/t_{c}$ values, both parameters remain nearly constant and show no significant variation with increasing $k_{B}T/t_{c}$. 
 
\subsection{Phase diagrams: case $|J_{K}|>|J_{H}|$}
\begin{figure}
    \centering    \includegraphics[scale=0.45]{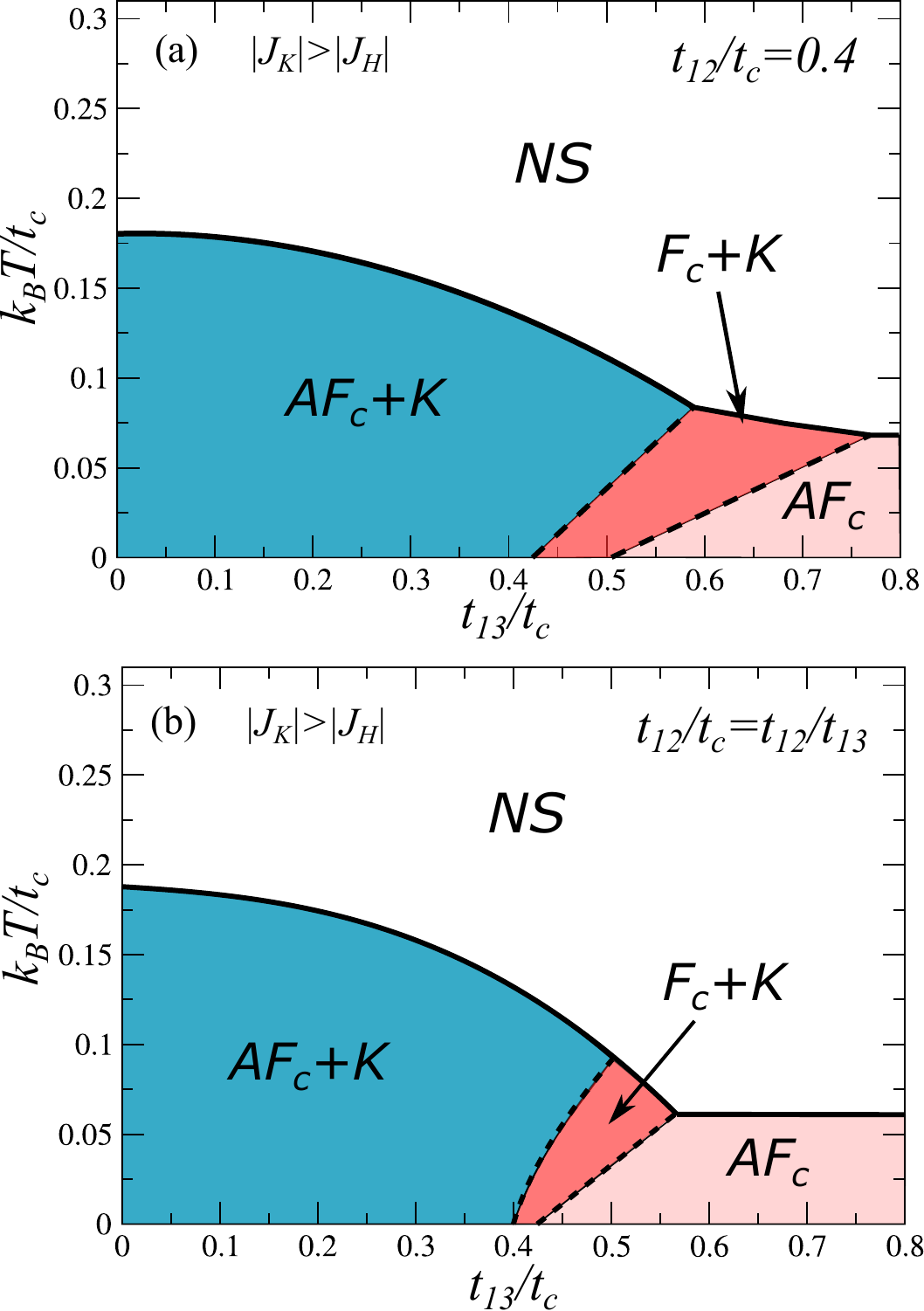}
   \caption{Phase diagrams of $k_{B}T/t_{c}$ versus $t_{13}/t_{c}$ with (a) $t_{12}/t_{c}=0.4$  and (b) $t_{12}/t_{c}=t_{13}/t_{c}$ in the case where the
Heisenberg interaction is large compared to the Kondo coupling:
$J_{K}=-0.5$, and $J_{H}=-0.1$. $W^{c}_{2(3)}=W_{d}=1$. $NS$ is a normal state, $AF_{c}$ are antiferromagnetic correlations, $F_{c}$ are ferromagnetic correlations and $K$ is a Kondo state. The continuous black lines are second-order transitions and the dashed black lines are firth-order transitions. }
    \label{DF2}
\end{figure}

The phase diagrams of $k_{B}T/t_{c}$ versus $t_{13}/t_{c}$ in the case where $|J_{K}| > |J_{H}|$ are shown in Fig. \ref{DF2}. In this case, \textit{layer 1} interacts with \textit{layer 2} with a finite value of $t_{12}/t_{c} = 0.4$. At low values of $t_{13}/t_{c}$ and as $k_{B}T/t_{c}$ decreases, the presence of a continuous transition (continuous black lines) from $NS$ to $AF_{c}+K$ is observed. At low values of $k_{B}T/t_{c}$ and as the interaction between \textit{layer 1} and \textit{layer 3} increases, discontinuous transition (dashed black lines) from $AF_{c}+K$ to $F_{c}+K$ is observed. Subsequently, with the increase of $t_{13}/t_{c}$, a new discontinuous transition from $F_{c}+K$ to $AF_{c}$ is shown. At high intensities of $t_{13}/t_{c}$ and with the increase of $k_{B}T/t_{c}$, a continuous transition between $AF_{c}$ and $NS$ is observed. Additionally, in Fig. \ref{DF2}(b) we show the phase diagram of $T$ versus $t_{13}/t_{c}$ in the case where $t_{12}/t_{c}= t_{13}/t_{c}$.  The behavior of the phase diagram is similar to the previous one, see Fig. \ref{DF1}(a), where at low intensities of $t_{13}/t_{c}$ and as $k_{B}T/t_{c}$ decreases, a continuous transition from $NS$ to $AF_{c}+K$ is shown. Also, for low values of $k_{B}T/t_{c}$ and as $t_{13}/t_{c}$ is increasing, a discontinuous transition from $AF_{c}+K$ to $AF_{c}+K$, followed by a discontinuous transition from $AF_{c}+K$ to $F_{c}+K$, to finally end in a new transition from $F_{c}+K$ to $AF_{c}$. 

\section{Conclusion and experimental evidence}\label{Section3}

In this work, we investigated the effects of interfacial proximity in a heterostructure displaying the Kondo effect and short-range magnetic correlations. Our heterostructure comprises three layers, where the first layer (\textit{layer 1}) is governed by KHl model involving $f$- and $d$- electrons interacting via Kondo and Heisenberg couplings. The other two layers (\textit{layer 2} and \textit{3}) consist of non-interacting itinerant $c$-electrons, where the coupling to KHl is defined by two perpendicular hopping terms, $t_{12}/t_{c}$ and $t_{13}/t_{c}$. Additionally, we have proposed two possible cases:  $|J_{K}|<|J_{H}|$ and $|J_{K}|>|J_{H}|$, at \textit{half-band filling}. We demonstrate that by varying $t_{12}/t_{c}$ and $t_{13}/t_{c}$, electronic dynamics are induced at the KHl interface, altering the behavior of the mean-field parameters describing the Kondo effect ($\lambda_{11}/t_{c}$) and short-range magnetic correlations ($\Gamma_{11}/t_{c}$). In the case where $|J_{K}|<|J_{H}|$, $AF_{c}$ are shown to dominate in the phase diagram of $k_{B}T/t_{c}$ versus $t_{12}/t_{c}$ when the distance between layers is symmetric, $t_{12}/t_{c}=t_{13}/t_{c}$. Additionally, a \textit{dome} is observed where $K$ coexist with $AF_{c}$ when the distances between layer interfaces are different,  i.e., $t_{12}/t_{c}\not=t_{13}/t_{c}$.
In the case where $|J_{K}|>|J_{H}|$, a sequence of discontinuous and continuous transitions is shown to evolve in the $k_{B}T/t_{c}$ versus $t_{13}/t_{c}$ phase diagrams, with the formation of regions of pure $AF_{c}$, as well as regions of coexistence of $AF_{c}+K$ and $F_{c}+K$. This case shows a rich phase diagrams with a mixture of $K$ and short-range magnetic correlations of different nature: $AF_{c}$ and $F_{c}$.
Through both cases, we propose that by alternatively studying the proximity effects between the interfaces in a Kondo-Heisenberg lattice and two metallic lattices, we can obtain Kondo and magnetic correlations, to those found in studies conducted outside of \textit{half-filling} regime in compounds such as \ce{CeCu6}, \ce{CeInCu2} and \ce{CeRu2Si2} \cite{Mignot1988, Regnault1988, Pierre1990, Liohneysen1998}. 

The approach carried out with heterostructures could be  an alternative route to study the \textit{Nozières exhaustion problem}. This classic problem arises when there are not enough conduction electrons in the lattice to screen all spins of localized electrons in a two-dimensional Kondo lattice \cite{Nozieres1998}.
We speculate that by varying the interlayer hopping terms of the Kondo heterostructure, we can induce tunneling of conduction electrons, between layers, capable of screening the localized spins in the KHl.

\section*{ACKNOWLEDGMENTS}
J. Faúndez, L. C. Prauchner and S. G Magalhaes thank the CNPq (Conselho Nacional de Desenvolvimento Científico e Tecnológico), processo: 200778/2022-6. J. Fáundez also acknowledges partial support from ANID Fondecyt grant number 3240320. S. E. Reyes-Lillo acknowledges support from ANID Fondecyt Regular grant number 1220986. Powered@NLHPC: This research was partially supported by the supercomputing infrastructure of the NLHPC (ECM-02). The authors would like to thank Miguel Gusmão for his valuable contribution during the discussions.
\newpage

\begin{widetext}
\appendix
\section{Green's Function}
\label{appendix}

The Greens functions obtained from the equation of motion method (Zubarev formalism \cite{Zubarev1960}) are given in the matrix representation as
\begin{align}
 - \braket{\braket{ \mathbf{ \Psi_{\textbf{k}\sigma}\otimes  \Psi^{\dagger}_{\textbf{k}\sigma}}  }}_{z}= (z-\mathbf{h_{\textbf{k}}})^{-1}.
   \label{A00}
\end{align}

Therefore, the minimum necessary set of Green's function is explicitly given as 

\begin{align}
 \braket{\braket{ f_{\textbf{k}1\sigma} :f^{\dagger}_{\textbf{k}1\sigma }  }}_{z}= 
- \frac{- t^2_{d1c3} \ (z-\epsilon_{c,2}(\textbf{k}))+[(z-\epsilon_{d,1}(\textbf{k})) \ (z-\epsilon_{c,2}(\textbf{k}))-t^{2}_{d1c2}] \ (z-\epsilon_{c,3}(\textbf{k}))}{  D_3(\textbf{k},z)}  \ ,
    \label{A7}
\end{align}

\begin{align}
 \braket{\braket{ d_{\textbf{k}1\sigma} :f^{\dagger}_{\textbf{k}1\sigma }  }}_{z}=- \frac{2J_{k}\lambda_{11} 
 \ (z-\epsilon_{c,2}(\textbf{k})) \ (z-\epsilon_{c,3}(\textbf{k})) }{  D_3(\textbf{k},z)}  \ ,
    \label{A8}
\end{align}
\begin{align}
 \braket{\braket{ d_{\textbf{k}1\sigma} :d^{\dagger}_{\textbf{k}1\sigma }  }}_{z}= -\frac{(z-B\epsilon_{d,1}(\textbf{k})-E_{0}) \ (z-\epsilon_{c,2}(\textbf{k})) \ (z-\epsilon_{c,3}(\textbf{k})}{  D_3(\textbf{k},z)}  \ ,
   \label{C8}
\end{align}

\begin{align}
 %\boxed{
 \braket{\braket{ c_{\textbf{k}2\sigma} :c^{\dagger}_{\textbf{k}2\sigma }  }}_{z}=- \frac{-t^2_{d1c3} \ (z-B\epsilon_{d,1}(\textbf{k})-E_0) + 
 [(z-B\epsilon_{d,1}(\textbf{k})-E_{0}) \ (z-\epsilon_{d,1}(\textbf{k})) - 4 J^2_K \lambda_{11}^2] \ (z-\epsilon_{c,3}(\textbf{k})) }{  D_3(\textbf{k}, z)}  \ ,
 %}
   \label{C9}
\end{align}
\begin{align}
 %\boxed{
 \braket{\braket{ c_{\textbf{k}3\sigma} :c^{\dagger}_{\textbf{k}3\sigma }  }}_{z}=- \frac{-t^2_{d1c2} \ (z-B\epsilon_{d,1}(\textbf{k})-E_0) + 
 [(z-B\epsilon_{d,1}(\textbf{k})-E_{0}) \ (z-\epsilon_{d,1}(\textbf{k})) - 4 J^2_K \lambda_{11}^2] \ (z-\epsilon_{c,2}(\textbf{k})) }{  D_3(\textbf{k}, z)}  \ ,
 %}
   \label{C91}
\end{align}

\begin{align}
 \braket{\braket{ d_{\textbf{k}1\sigma} : c^{\dagger}_{\textbf{k}2\sigma }}}_{z} =-\frac{(z-\epsilon_{c,2}(\textbf{k})) \ (z-B\epsilon_{d,1}(\textbf{k})-E_0) \ t_{d1c2}}{D_3(\textbf{k},\omega)}  \ ,   
 \label{C10}
 \end{align}
\begin{align}
 \braket{\braket{ d_{\textbf{k}1\sigma} : c^{\dagger}_{\textbf{k}3\sigma }}}_{z} =-\frac{(z-\epsilon_{c,3}(\textbf{k})) \ (z-B\epsilon_{d,1}(\textbf{k})-E_0) \ t_{d1c3}}{D_3(\textbf{k},\omega)}  \ ,
   \label{C11}
\end{align}

where

\begin{align}
    D_3(\textbf{k},z) =
     [(z-\epsilon_{d,1}(\textbf{k})) \ (z-\epsilon_{c,2}(\textbf{k})) \ (z-\epsilon_{c,3}(\textbf{k}))-\nonumber\\t_{d1c2}^{2} \  (z-\epsilon_{c,3}(\textbf{k}) - t_{d1c3}^{2} (z-\epsilon_{c,2}(\textbf{k})] \ (z-B\epsilon_{d,1}(\textbf{k})-E_{0}) 
    & - 4J^{2}_{k}\lambda_{11}^{2}(z-\epsilon_{c,2}(\textbf{k}))\ (z-\epsilon_{c,3}(\textbf{k})).
\label{D3}
\end{align}
\end{widetext}

%-----------------------------------------------------------------------------------------------%
%------------------------------------------------------------------------------------------------%

%%%%%%%%%%%%%%%%%%%%%%%%%%%%%%%%%%%

\end{document}